\documentclass[12pt]{article}
\input{epsf}
\usepackage{graphicx}
\def\AJ{{\it Astroph. J.} }
\def\AJL{{\it Ap. J. Lett.} }

\def\ASJ{{\it Astron. J.} }

\def\CQG{{\it Class. Quantum Gravity} }

\def\FP{{\it Fortschr. Physik} }
\def\GRG{{\it Gen. Relativity and Gravitation} }

\def\JHEP{{\it JHEP} }

\def\MPL{{\it Mod. Phys. Lett.} }

\def\NAT{{\it Nature} }
 
\def\NC{{\it Il Nuovo Cimento} }
\def\NP{{\it Nucl. Phys.} }
\def\PL{{\it Phys. Lett.} }
\def\PR{{\it Phys. Rev.} }
\def\PRL{{\it Phys. Rev. Lett.} }

\def\frac#1#2{{\textstyle{{#1}\over {#2}}}}

\def\lsim{\mathrel{\rlap{\lower4pt\hbox{\hskip1pt$\sim$}}
    \raise1pt\hbox{$<$}}}
\def\gsim{\mathrel{\rlap{\lower4pt\hbox{\hskip1pt$\sim$}}
    \raise1pt\hbox{$>$}}}
\def\sqr#1#2{{\vcenter{\vbox{\hrule height.#2pt
         \hbox{\vrule width.#2pt height#1pt \kern#1pt
         \vrule width.#2pt}
         \hrule height.#2pt}}}}

\def\beq{\begin{equation}}
\def\eeq{\end{equation}}
\def\beqa{\begin{eqnarray}} 
\def\eeqa{\end{eqnarray}}

\def\laq{\raise 0.4 ex \hbox{$<$}\kern -0.8 em\lower 0.62 ex\hbox{$\sim$}}
\def\gaq{\raise 0.4 ex \hbox{$>$}\kern -0.7 em\lower 0.62 ex\hbox{$\sim$}}

\begin{document}

\begin{center}
{\large \bf WMAP Constraints on the Generalized Chaplygin
 Gas Model }
\vspace{0.5cm}

M. C. Bento\footnote{Also at CFIF, Instituto Superior T\'ecnico, Lisboa.
  Email address: bento@sirius.ist.utl.pt},
O. Bertolami 
\footnote{Also at CFNUL, Universidade de Lisboa.
 Email address: orfeu@cosmos.ist.utl.pt }
and A. A. Sen
\footnote{Also at CENTRA, Instituto Superior T\'ecnico, Lisboa.
 Email address: anjan@x9.ist.utl.pt}
\vspace{0.5cm}

{ Departamento de F\'\i sica, Instituto Superior T\'ecnico \\
Av. Rovisco Pais 1, 1049-001 Lisboa, Portugal}

\vspace{0.5cm}
\centerline{May 5, 2003}
\vspace{0.5cm}
\end{center}

\centerline{\bf Abstract}
\vspace{0.3cm}

The generalized Chaplygin gas (GCG) model explains the recent 
accelerated expansion of the Universe via an exotic background fluid 
whose equation of state is given by
$p = - A/\rho^{\alpha}$, where $A$ 
is a positive constant and $0 < \alpha \le 1$. The model  is an interesting 
alternative 
to  scenarios  involving scalar field potentials, with the ensuing 
unnatural fine tuning conditions for the underlying particle physics theories. 
We derive constraints on the parameter space of the  model from bounds on 
the location of the first few peaks and troughs of the the Cosmic Microwave
Background Radiation (CMBR) power spectrum arising from recent WMAP and 
BOOMERanG data.

\vspace{1cm}

 PACS number(s): 98.80.Cq


\section{Introduction}

It has recently been proposed that the  evidence for a dark energy 
component to the total energy density of  the Universe at 
present  might be explained by a change in the equation 
of state of the background fluid  rather than by  a cosmological constant 
or the dynamics of a scalar field rolling down a potential \cite{Kamenshchik}. 
This allows, at least in principle, to avoid well known fine-tuning problems 
associated with $\Lambda$CDM and  quintessence models.
Within the framework of 
Friedmann-Robertson-Walker cosmology, one considers   
an exotic background fluid, the GCG, which is 
described by the following equation of state

\beq
p_{ch} = - {A \over \rho_{ch}^\alpha}~~,
\label{eq:eqstate}
\eeq
\vskip 0.3cm

\noindent
where $\alpha$ is a constant in the range 
$0 < \alpha \le 1$ (the Chaplygin gas corresponds to the case $\alpha=1$) and
$A$ a positive constant. Inserting this equation of state 
into the relativistic energy conservation equation, leads to a density evolving
as \cite{Bento1}

\beq
\rho_{ch} =  \left(A + {B \over a^{3 (1 + \alpha)}}\right)^{1 \over 1 +
 \alpha}~~,
\label{eq:rhoc}
\eeq 
\vskip 0.3cm

\noindent
where $a$ is the scale-factor of the Universe and $B$ an integration 
constant. It is remarkable that the model interpolates between 
a universe dominated by dust and a De Sitter one via an intermediate
phase which is a mixture of a cosmological constant and a perfect
fluid with a ``soft''
matter equation of state, $p = \alpha \rho$ ($\alpha \not= 1$) \cite{Bento2}.
Notice that even though Eq. (\ref{eq:eqstate}) admits a wider range of
positive $\alpha$ values, the chosen range  ensures that 
the sound velocity ($c_s^2 = \alpha A/ \rho_{ch}^{1+\alpha}$) does not exceed,
in the ``soft'' equation of state phase, 
the velocity of light. Furthermore, as pointed out in Ref. \cite{Bento1}, 
it is only for $0 < \alpha \le 1$ 
that the analysis of the evolution of energy density fluctuations is
 meaningful.
  
Furthermore, as discussed in Ref. \cite{Bento1}, the model can be described 
by a
complex scalar field whose action can be written as a generalized Born-Infeld
action. Recently, it has been shown that a curvature self-interaction 
of the cosmic gas can mimic the GCG equation of state \cite{Schwarz}.
It is quite clear that the GCG is a candidate for explaining the observed
accelerated expansion of the Universe \cite{Perlmutter} as it automatically 
leads to an asymptotic phase where the equation of state is dominated by a 
cosmological constant, $8 \pi G A^{1/1+\alpha}$. 
It has also been shown that the model admits, under conditions, 
an inhomogeneous generalization which can be regarded as a unification 
of dark matter and dark energy \cite{Bento1,Bilic} without conflict with 
standard structure formation
scenarios \cite{Bento1,Bilic,Fabris}.  Hence, it is fair to conclude that 
the GCG model is  an interesting alternative to models where the 
accelerated
expansion of the Universe is 
explained via  an uncancelled cosmological constant (see \cite{Bento2} and 
references therein) or a  scalar field potential as in quintessence 
models with one \cite{quint1} or two scalar fields \cite{quint2}.
Recently, some questions have been raised concerning the viability of the 
GCG model. For instance, in Ref.~\cite{Tegmark}, it is  claimed that
the model produces a  matter power
spectrum inconsistent with observation; however, the authors did not 
include the effect of baryons, 
which should  play a crucial role and, in particular, would 
require a two-fluid analysis, as was done in Ref.~\cite{Beca}, with 
the conclusion that the GCG can be quite different from the $\Lambda$CDM 
model and still reproduce 2dF large scale strucure data. On the other 
hand, in Ref. 
\cite{Bean}, it is argued that the GCG model is indistinguishible from the
$\Lambda$CDM model, which is not surprising as the authors did  not consider 
the GCG  as  an entangled mixture of dark matter and dark energy as 
expected in a unification model.

The possibility of describing dark energy via the GCG model 
has led to a wave of interest aiming 
to constrain the model using  observational data, 
particularly those arising from SNe Ia 
\cite{Supern,Makler,Alcaniz} and gravitational lensing statistics 
\cite{Silva}.

In this work, we  extend the analysis carried out in Ref. \cite{Bento4} 
(see also Ref.~\cite{Carturan} for a study based on the CMBfast code) aiming 
to constrain the parameters of the GCG model from recent bounds 
on the positions of peaks and troughs of 
the CMBR power spectrum,  employing basically the same  methods that 
have been used  
to constrain quintessence models (see e.g. 
Refs.~\cite{Doran1,Doran2,Domenico,Barreiro}). Restricting the 
analysis of the CMBR power spectrum to the locations of peaks and 
troughs rather than considering the structure of the whole spectrum 
turns out to be a simple but very powerful tool in constraining the 
model parameters basically because of the precision with which these 
positions are now determined, especially following WMAP results.
We find that the model is 
compatible with WMAP bounds on the locations of  the first two peaks and 
first trough, and
BOOMERanG bounds on the location of  the third peak provided 
$\alpha\lsim 0.6$, thus ruling out  the Chaplygin gas model. The allowed 
range of model parameters depends, in particular, on $h$ and $n_s$; for 
instance, for $h=0.71$ and $n_s=1$, we obtain $\alpha\lsim 0.4,~
0.76\lsim A_s\lsim 0.88$. These bounds become tighter for $n_s<1$ e.g. for 
$n_s=0.93$, we get $\alpha\lsim 0.2,~0.79\lsim A_s\lsim 0.82$. The allowed 
regions of model parameters become slightly larger for smaller values of $h$.

Finally, we should like to mention that, in order to make the Chaplygin gas 
model consistent with the location of peaks and troughs in the CMBR
power spectrum as measured by WMAP, values of 
$h$ smaller than the ones suggested by WMAP data are required, namely $h\lsim 
0.65$,
together with the condition that $n_s$  is close to $1$.

\section{CMBR constraints for the GCG model}

The CMBR peaks arise from acoustic oscillations of the primeval plasma just
before the 
Universe becomes transparent. The angular momentum scale of the oscillations
is set by the acoustic scale, $l_A$, which for a flat Universe is given by

\beq
\label{eq:la}
l_A = \pi {\tau_0 - \tau_{\rm ls} \over \bar c_s \tau_{\rm ls}}~~,
\eeq
\vskip 0.3cm

\noindent
where $\tau = \int a^{-1} dt$ is the conformal time, $\tau_0$ and 
$\tau_{\rm ls}$ being  its value  today and at
last scattering respectively, while $\bar{c}_s$ is the average sound speed 
before decoupling:

\beq
{\bar c}_s \equiv \tau_{ls}^{-1} \int_0^{\tau_{ls}} c_s\ d\tau~,
\label{eq:bcs}
\eeq
where

\beq
c_s^{-2} = 3+{9\over 4}{\rho_b(t)\over\rho_\gamma(t)}~,
\label{eq:cs1}
\eeq
with $\rho_b/\rho_\gamma$ the ratio of baryon to photon energy density.

In an idealised model of the primeval plasma, there is a simple relationship
between the location of the $m$-th peak and the acoustic scale, namely
$l_m\approx m l_A$. However, the  peaks position is shifted
by several effects which can be estimated by parametrising the location of 
the $m$-th peak, $l_m$, as in \cite{Doran1,Hu}

\beq 
\ell_{p_m} \equiv \ell_A \left(m - \varphi_m\right)
\equiv \ell_A (m -\bar \varphi-\delta \varphi_m)~, 
\label{eq:lm}
\eeq
where $\bar \varphi\equiv \varphi_1$ is the overall peak shift and
$\delta \varphi_m\equiv \varphi_m-\bar \varphi$ is the relative shift of the
m-th peak relative to the first one. Eq.~(\ref{eq:lm}) can also be used for
the position of troughs if one sets, $m=3/2$ for the first trough and $m=5/2$ 
for the second trough. Even though analytical
relationships between the cosmological parameters and the peak shifts are not 
available,
one can use fitting formulae describing their dependence on these
parameters. We use the  formulae given in Ref. \cite{Doran2} for the first 
three peaks and first trough,  which we reproduce in the Appendix, for 
convenience. It is relevant pointing out that although these formulae
were obtained for quintessence models with an exponential potential, they 
are expected 
to be fairly independent of the form of the potential and the nature of 
the late time acceleration mechanism  
as the shifts are practically independent of post recombination
physics. We
should stress that the analytic estimators we are using, determined by
comparison with CMBfast for standard models, is less than one percent
\cite{Doran2}.

Following our dark matter-energy
unification scenario, we rewrite the energy density, Eq. (\ref{eq:rhoc}), as

\beq
\label{eq:rho_c0}
\rho_{ch}= \rho_{ch0}\left( A_s + {(1-A_s)\over
a^{3(1+\alpha)}}\right)^{1/1+\alpha}~~,
\eeq
where $A_s\equiv A/ \rho_{ch0}^{1+\alpha}$ and
$\rho_{ch0}=(A+B)^{1/ 1+\alpha}$. In terms of the new variables, 
Friedmann equation reads

\beq
\label{eq:H2}
H^2  =  {8\pi G\over 3}\left[{\rho_{r0}\over a^4}+{\rho_{b0}\over{a^3}} 
+ \rho_{ch0} \left( A_s + 
      {(1-A_s)\over a^{3(1+\alpha)}}\right)^{1/1+\alpha}\right]~~,
\eeq
where we have included the contribution of radiation and baryons as these are 
not accounted for by the GCG equation of state. 

Several important features of  Eq. (\ref{eq:rho_c0}) are worth remarking. 
Firstly, we mention that $A_s$ must lie in the interval $0 \leq A_{s} \leq 1$ 
as otherwise $p_{ch}$ would be undefined for some value of the scale-factor. 
Secondly, for $A_{s} = 0$,
the Chaplygin gas behaves as  dust and, for $A_{s} = 1$, it behaves like a 
cosmological constant. Notice that   the Chaplygin gas
corresponds to a $\Lambda$CDM model only  for $\alpha = 0$ . 
Hence, for the chosen range of $\alpha$, the GCG model is 
clearly different from the $\Lambda$CDM model. Another relevant issue is that
the sound velocity of the fluid is given, at present, by $\alpha A_s$ and
thus $\alpha A_s \le 1$. Using the fact that 
$\rho_{r0}/\rho_{ch0}=
\Omega_{r0}/ (1-\Omega_{r0}-\Omega_{b0})$
and
$\rho_{b0}/ \rho_{ch0}=
\Omega_{b0}/ (1-\Omega_{r0}-\Omega_{b0})$,
we obtain

\beq
\label{eq:H2final}
H^2=\Omega_{ch0} H_0^2 a^{-4} X^2(a)~~,
\eeq
with

\beqa
\label{eq:defX}
X(a)&  = & {\Omega_{r0}\over 1-\Omega_{r0}-\Omega_{b0}} + 
             {\Omega_{b0}~a \over 1-\Omega_{r0}-\Omega_{b0}} \nonumber\\
& + & a^4 \left( A_s + {(1-A_s)\over
a^{3(1+\alpha)}}\right)^{1/1+\alpha}~~.
\eeqa
Moreover, since $H^2=a^{-4} \left(d a\over d \tau\right)^2$, we get

\beq
\label{eq:dtau}
d\tau={da\over \Omega_{ch0}^{1/2} H_0 X(a)}~~,
\eeq
so that

\beq
\label{eq:lA}
l_A={\pi\over \bar c_s}\left[{\int_0}^1 {da\over X(a)} \left( {\int_0}^{a_{ls}}
{da\over X(a)}\right)^{-1} - 1 \right]~~.
\eeq
where $a_{ls}$ is the scale factor at last scattering, for which we use the 
fitting formula \cite{Hu1996}

\beq
a_{ls}^{-1}-1=z_{ls}=1048[1+0.00124 w_b^{-0.738}][1+g_1 w_m^{g_2}]~,
\label{eq:zls}
\eeq
where

\beqa
g_1 &=& 0.0783 w_b^{-0.238} [1+39.5 w_b^{0.763}]^{-1}~,\nonumber\\
g_2 &=& 0.56 [1+21.1 w_b^{1.81}]^{-1}~,
\label{eq:coef}
\eeqa
and  $\omega_{b,m}\equiv \Omega_{b,m} h^2$.

Let us now turn to the discussion of the available CMBR data.
The bounds on the locations of the first two  peaks and the first trough,
from  WMAP measurements of the CMBR temperature angular power spectrum 
\cite{WMAP}, are

\beqa
\ell_{p_1} &=& 220.1\pm 0.8~,\nonumber\\  
\ell_{p_2} &=& 546\pm 10~,\nonumber\\  
\ell_{d_1} &=& 411.7\pm 3.5~;   
\label{eq:wmap}
\eeqa
where all uncertainties are 1$\sigma$ and include calibration and
beam errors.
The location of the third peak, from  BOOMERanG measurements, is given by 
\cite{Boomerang}

\beq
\ell_{p_3} = 825^{+10}_{-13}~.
\label{eq:l3}
\eeq

 From the computation of the acoustic scale,
Eq. (\ref{eq:lA}), the equation for the shift of the peaks, Eq.~(\ref{eq:lm}), 
and the fitting  formulae given  in the Appendix, we  look for the 
combination of GCG model 
parameters that is consistent with observational bounds.  
Our results  are shown in 
Figures 1 - 4, where  have assumed that $\omega_b=0.0224$ and used the 
fact that  $A_s$ and $\Omega_m$ are related by

\beq
A_s = {1 - \Omega_{m}-\Omega_{r} \over 1 -\Omega_{b}-\Omega_{r}}~~.
\label{eq:AOmega}
\eeq
which is obtained by noting that 
for $\alpha = 0$ the model is just the $\Lambda$CDM model; thus, one 
should identify the Chaplygin gas parameters with the usual
density parameters when substituing $\alpha = 0$ in 
Eq. (8) (for the  present, $a_0 = 1$), taking into account that 
$\Omega_m = \Omega_b + \Omega_{CDM}$.

In Fig. 1, we  plot contours in the $(h,\Omega_m)$ plane corresponding 
to the bounds on the 
first three peaks and first trough of the CMBR power spectrum, 
Eqs.~(\ref{eq:wmap}) - (\ref{eq:l3}), for 
$n_s=0.97$  and different values of $\alpha$. The box on the 
$\alpha=0$ ($\Lambda$CDM model)  plot corresponds to the bounds on 
$h$ and $\omega_m$  arising
from the combination of WMAP data with other CMB experiments (ACBAR and CBI),
2dFGRS measurements and Lyman $\alpha$ forest  data  \cite{WMAP}:

\beq
h=0.71^{+0.04}_{-0.03}~,~~~\omega_m =0.135^{+0.008}_{-0.009}~.
\label{eq:wmapc}
\eeq
Notice that the above bound on $h$ is slightly more restrictive than
the bound obtained from WMAP data alone \cite{WMAP}

\beq
h=0.72\pm 0.05~.
\label{eq:wmaponly}
\eeq

Figures 2 and 3 show the same contours but for $n_s=1$ and $1.03$, 
respectively. In Figs. 4 and 5, contours are shown in the $(A_s,\alpha)$
plane for $h=0.71$ and $h=0.6$.

\section{Discussion and Conclusions}

In this work we have shown that current bounds on the  location of the first 
few peaks and troughs in the CMBR power spectrum,
as determined from WMAP and BOOMERanG data, allow constraining a 
sizeable portion of the parameter space of the GCG model.
Our results indicate that WMAP bounds imply that the 
Chaplygin gas model ($\alpha= 1$ case)  is ruled out and so are 
models with $\alpha > 0.6$. For low values of $n_s$,  
$\alpha = 0.6$ is also ruled out. 
However, for  $n_s>0.97$, $\alpha = 0.6$ becomes increasingly 
compatible with data. Hence, one can 
safely state that models with $\alpha \le 0.2$ are always consistent. 

Our analysis shows that results depend strongly on
the Hubble parameter and since WMAP's bound on this quantity
 was obtained for $\Lambda$CDM
models and, on the other hand, 
 there are recent determinations of the  Hubble constant,
 combining Sunayev-Zeldovich and X-ray flux measurements of clusters
 of galaxies, that give much lower values of $H_0$, namely  \cite{Reese}

\beq
H_0=60\pm 4 ^{+13}_{-18}~\mbox{km/s/Mpc}~,
\label{eq:reese}
\eeq
and \cite{Mason}

\beq
H_0=66^{+14}_{-11}\pm 15~\mbox{km/s/Mpc}~,
\label{eq:Mason}
\eeq
it is relevant to examine the implications, in particular regarding 
the exclusion of the  Chaplygin gas model, of  relaxing the bound
(\ref{eq:wmaponly})  and allow for lower values of $h$. 
Figures 4 and 5 show that, for $h = 0.71$ (the central value 
for WMAP's bound on $h$), $\alpha = 1$ is not allowed for any combination 
of parameters; however, for $h = 0.6$  (slightly below WMAP's preferred 
range), $\alpha = 1$ is allowed provided $n_s$ is around $1$. 
In fact, a deeper analysis shows that, in order for  the Chaplygin gas model 
to become consistent with peak and dip locations of the CMBR power
spectrum, it is necessary that $h\lsim 0.65$ and $n_s\approx 1$. 

These results are compatible with the ones 
found in Ref. \cite{Bento4} using bounds on  the 
third peak from  BOMERanG and the first peak from 
Archeops \cite{Benoit} data  as well as bounds from  SNe Ia  and distant 
quasar sources, 
namely $0.2\lsim \alpha \lsim 0.6$ and $0.81\lsim A_s \lsim 0.85$. 
We find,in particular, that bounds from SNe Ia data, which suggest that
$0.6 \lsim A_s \lsim 0.85$ \cite{Makler}, are consistent with our present 
results for $n_s=1$ and $h=0.71$, namely $0.78\lsim A_s \lsim 0.87$. 

\vspace{0.5cm}

{\it Note added } After we had completed this work,  a related study
has appeared \cite{Amendola} which makes a likelihood analysis based 
on the full WMAP CMB
data set using  a modified CMBfast code, with results similar to ours.

\vspace{0.5cm}

\centerline{\bf {Acknowledgments}}

\noindent
M.C.B. and  O.B.
acknowledge the partial support of Funda\c c\~ao para a 
Ci\^encia e a Tecnologia (Portugal)
under the grant POCTI/1999/FIS/36285. The work of A.A.S. is fully 
financed by the same grant. 

\vfill
\newpage
\centerline{\bf Appendix}

\vskip 0.5cm

We reproduce here the analytic approximations for the phase shifts found in
Ref.~\cite{Doran2}.
The overall phase shift is given by
\beq
{\bar\varphi}=(1.466-0.466 n_s)\left[a_1 r_*^{a_2}
+ 0.291 {\bar\Omega}_{ch}^{ls} \right]~,
\eeq   
where

\beqa
a_1 &=& 0.286+0.626\omega_b\nonumber\\
a_2 &=& 0.1786-6.308\omega_b+174.9\omega_b^2-1168\omega_b^3
\label{eq:as}
\eeqa
 are fitting coefficients,

\beq
{\bar \Omega}_{ch}^{ls}=\tau_{ls}^{-1} \int_0^{\tau_{ls}} \Omega_{ch}(\tau)
 d\tau~,
\label{eq:baro}
\eeq
and

\beq
r_*\equiv \rho_{rad}(z_{ls})/\rho_{m}(z_{ls})
\eeq
is the ratio of radiation to matter at decoupling and  $z_{ls}$ is given by
 Eqs. (\ref{eq:zls}) - (\ref{eq:coef}).

There is no relative shift of the first acoustic peak, $\delta\varphi_1=0$,
and the relative shifts for the second and third peaks are given by

\beq
\delta\varphi_2=c_0-c_1r_*-c_2r_*^{-c_3}+0.05(n_s-1)~,
\label{eq:delphi2}
\eeq
where
\beqa
c_0 &=& -0.1+\left(0.213-0.123{\bar\Omega}_{ls}^{ch}\right)\nonumber\\
        &&\times\exp\left\{-\left(52-63.6 {\bar\Omega}_{ls}^{ch}\right)
          \omega_b\right\},\nonumber\\
c_1 &=& 0.015+0.063\exp\left(-3500\omega_b^2\right), \nonumber\\
c_2 &=& 6\times 10^{-6}+0.137(\omega_b-0.07)^2,\nonumber\\
c_3 &=& 0.8+ 2.3 {\bar\Omega}_{ls}^{ch}
+\left(70-126{\bar\Omega}_{ls}^{ch}\right)\omega_b~,
\label{eq:cs2}
\eeqa
and 

\beq
\delta\varphi_3=10-d_1r_*^{d_2}+0.08(n_s-1)~,
\eeq
with
\beqa
d_1 &=& 9.97+\left(3.3-3{\bar\Omega}_{ls}^{ch}\right)\omega_b,\nonumber\\
d_2 &=& 0.0016-0.0067{\bar\Omega}_{ls}^{ch} +
\left(0.196-0.22{\bar\Omega}_{ls}^{ch}\right)\omega_b\nonumber\\
&& + \left(2.25+2.77{\bar\Omega}_{ls}^{ch}\right)\times  10^{-5}\omega_b^{-1}.
\label{eq:ds}
\eeqa

The relative shift of the first trough is given by

\beq
\delta\varphi_{3/2}=b_0+b_1r_*^{1/3}\exp(b_2r_*)+0.158(n_s-1)~
\label{eq:delphi}
\eeq
with

\beqa
b_0 &=& -0.086-0.079 {\bar\Omega}_{ch}^{ls}
       -\left( 2.22-18.1{\bar\Omega}^{ls}_{ch} \right) \omega_b\nonumber\\
    &&-\left(140+403{\bar\Omega}^{ls}_{ch} \right)\omega_b^2~,\nonumber\\
b_1 &=& 0.39-0.98{\bar\Omega}^{ls}_{ch}
       -\left(18.1-29.2{\bar\Omega}_{ls}^{ch}\right)\omega_b\nonumber\\
    && +440\omega_b^2,\\
b_2 &=& -0.57-3.8\exp({-2365\omega_b^2})~,
\label{eq:bs}
\eeqa

\vfill
\newpage

\newpage

\centerline{\bf Figures}
\begin{figure*}[ht!]
\begin{center}
 \includegraphics[height=13cm]{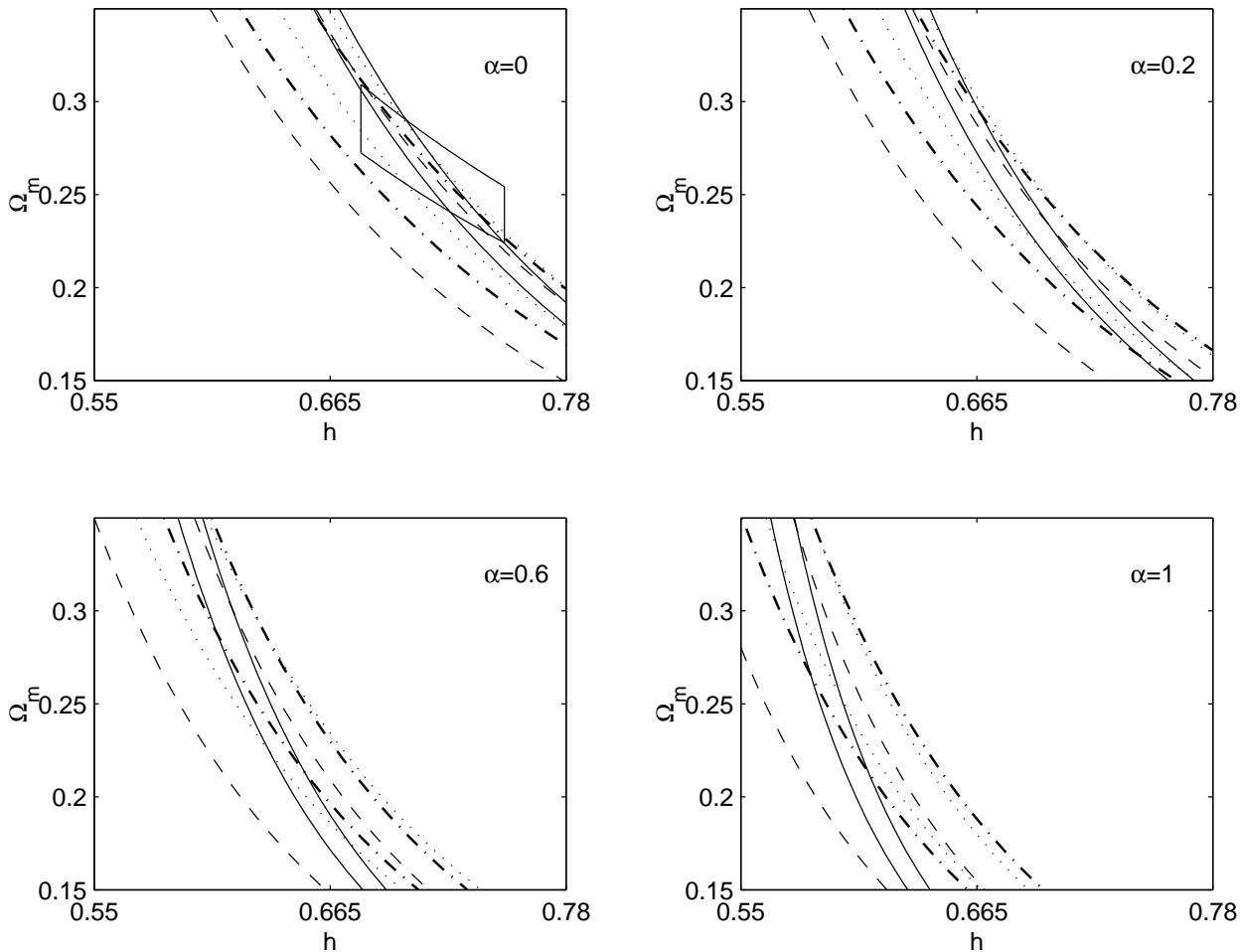}
 \caption{\label{fig:a0} Contour plots of the first three 
  Doppler peaks and first trough locations  in the
 $(\Omega_m, h)$ plane for GCG model,
 with $n_s=0.97$, for different values of 
 $\alpha$.
 Full, dashed, dot-dashed and dotted contours correspond to observational
 bounds on, respectively,
 $\ell_{p_1}$, $\ell_{p_2}$, $\ell_{p_3}$ and $\ell_{d_1}$, see
 Eqs.~(\ref{eq:wmap}) and (\ref{eq:l3}). The box on the $\alpha=0$ plot 
(corresponds to $\Lambda$CDM model)
indicates   the bounds on  $h$ and $\Omega_m h^2$ from  a combination of 
WMAP and
other experiments,  Eq.~(\ref{eq:wmapc}). }
\end{center}
\end{figure*}


\begin{figure*}[ht!]
\begin{center}
 \includegraphics[height=13cm]{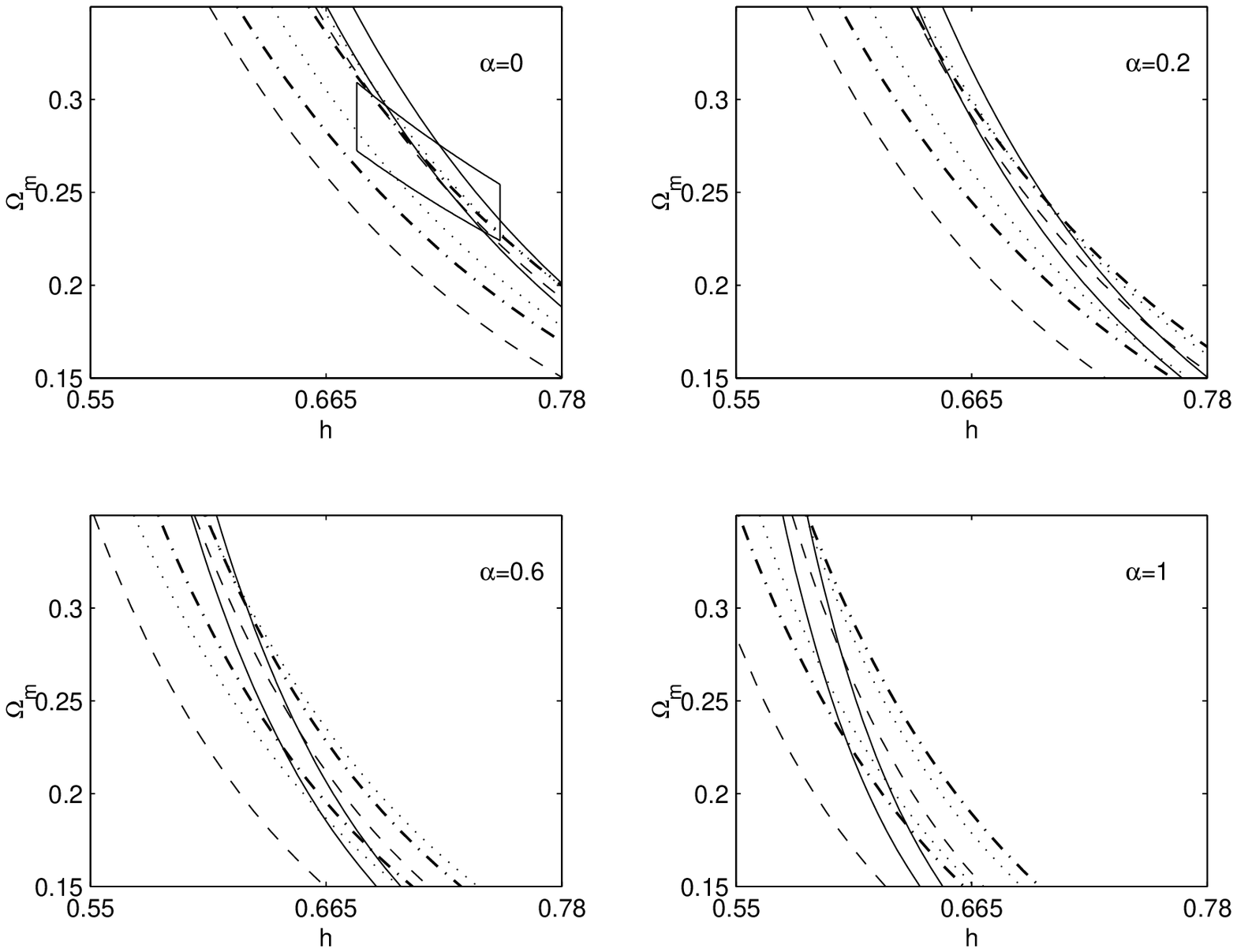}
 \caption{\label{fig:a03} As for Fig. 1 but with $n_s=1.0$.}
\end{center}
\end{figure*}

\begin{figure*}[ht!]
\begin{center}
 \includegraphics[height=13cm]{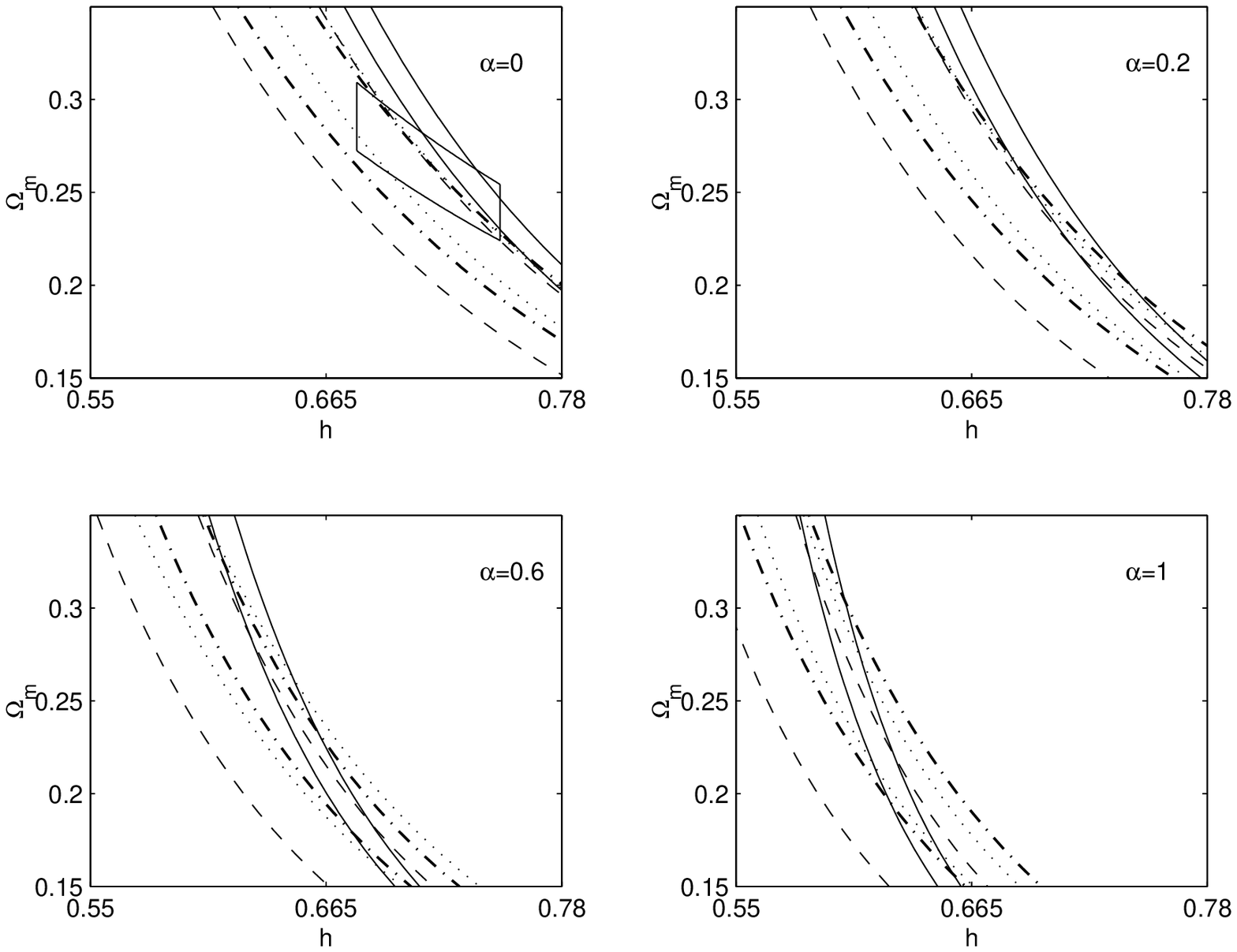}
 \caption{\label{fig:a06} As for Fig. 1 but with $n_s=1.03$.}
\end{center}
\end{figure*}

\begin{figure*}[ht!]
\begin{center}
 \includegraphics[height=13cm]{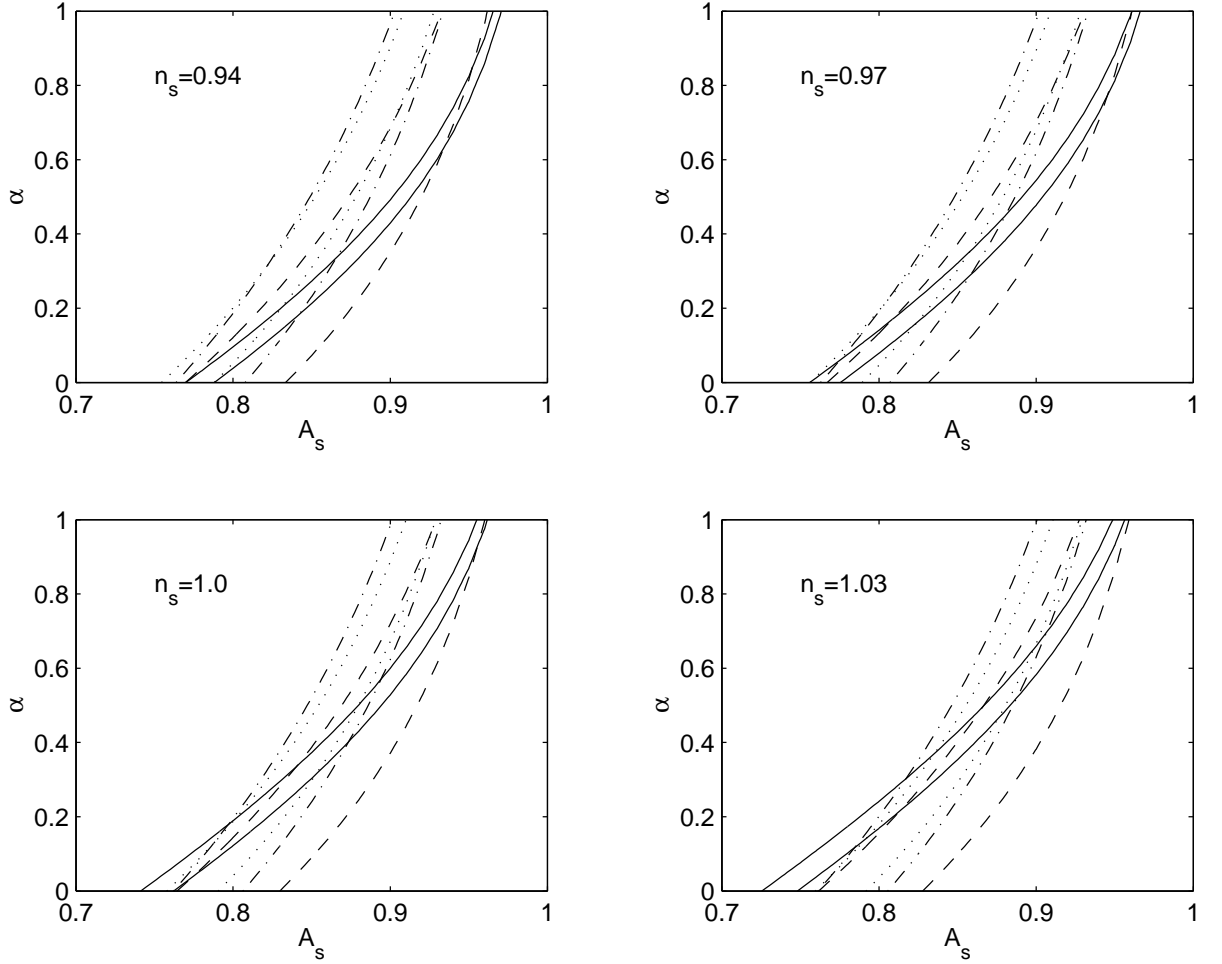}
 \caption{\label{fig:h71} Contour plots of the first three 
  Doppler peaks and first trough locations  in the
 $(A_s, \alpha)$ plane for GCG model,
 with $h=0.71$, for different values of 
 $n_s$.
 Full, dashed, dot-dashed and dotted contours correspond to observational
 bounds on,
 respectively,
 $\ell_{p_1}$, $\ell_{p_2}$, $\ell_{p_3}$ and $\ell_{d_1}$, see
 Eqs.~(\ref{eq:wmap}) and (\ref{eq:l3}).}
 \end{center}
\end{figure*}

\begin{figure*}[ht!]
\begin{center}
 \includegraphics[height=13cm]{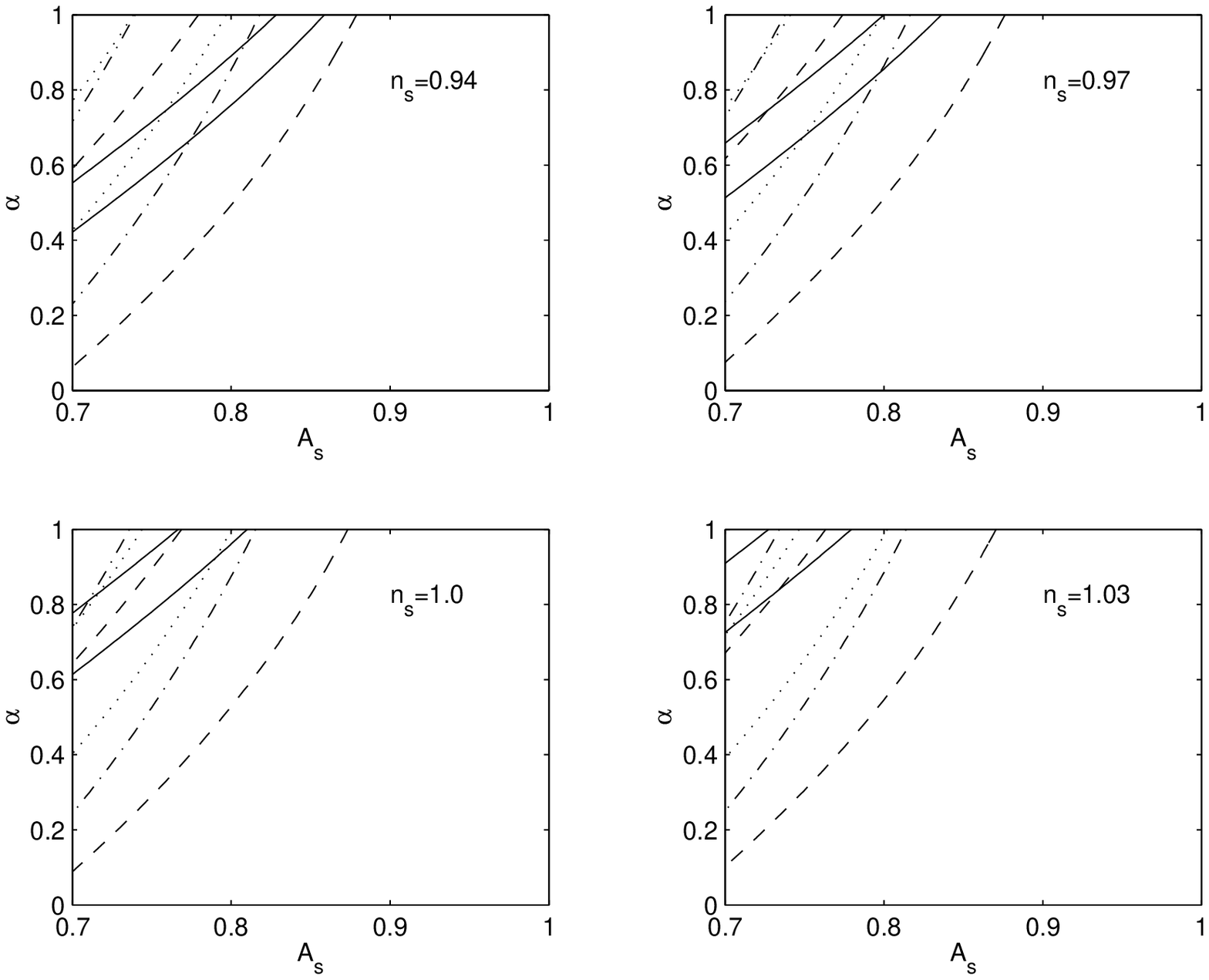}
 \caption{\label{fig:h6} As for Fig. 4 but 
 with $h=0.6$.}
 \end{center}
\end{figure*}


\begin{thebibliography}{99}



\bibitem{Kamenshchik} A. Kamenshchik, U. Moschella, V. Pasquier, \PL {\bf 511}
(2001) 265.


\bibitem{Bento1} M.C. Bento, O. Bertolami, A. A. Sen, \PR {\bf D66} (2002)
 043507.

\bibitem{Schwarz}  A.B. Balakin, D. Pav\'on, D.J.  Schwarz, W. Zimdahl,
 astro-ph/0302150.

\bibitem{Perlmutter} S.J. Perlmutter  et al.  (Supernova Cosmology
Project), \AJ 483(1997)565; \NAT {\bf 391} (1998) 51;
A.G.  Riess  et al., (Supernova Search Team) \ASJ {\bf 116} (1998) 1009;
P.M. Garnavich  et al., \AJ {\bf 509} (1998) 74.


\bibitem{Bilic} N. Bili\'c, G.B. Tupper, R.D. Viollier, \PL {\bf B535}
 (2002) 17. 


\bibitem{Fabris} J.C. Fabris, S.B.V. Gon\c calves, P.E. de Souza, 
\GRG {\bf 34} (2002) 53.


\bibitem{Bento2} M.C. Bento, O. Bertolami,  \GRG {\bf  31} (1999)
1461; M.C. Bento, O. Bertolami, P.T. Silva, \PL {\bf B498} (2001) 62.


\bibitem{quint1} M. Bronstein, {\it Phys. Zeit. Sowejt Union} {\bf 3} 
(1933) 73; 
O.  Bertolami, \NC  {\bf 93B} (1986) 36; \FP {\bf
34} (1986) 829; 
M. Ozer, M.O. Taha, \NP {\bf B287} (1987) 776;
 B. Ratra, P.J.E.  Peebles,  \PR {\bf  D37} (1988)
3406; \AJL {\bf 325} (1988) 117;
  C. Wetterich, \NP {\bf B302} (1988) 668;
 R.R.  Caldwell, R.  Dave, P.J.  Steinhardt, \PRL
{\bf 80} (1998) 1582;
  P.G.  Ferreira, M.  Joyce, \PR {\bf  D58} (1998)
023503; I. Zlatev, L. Wang, P.J. Steinhardt, \PRL {\bf 82}
(1999) 986; P. Bin\'etruy, \PR {\bf D60} (1999) 063502;
J.E. Kim, \JHEP 9905 (1999) 022;
 J.P. Uzan, \PR {\bf D59} (1999) 123510; 
T. Chiba, \PR {\bf D60} (1999) 083508;
L. Amendola, \PR {\bf D60} (1999) 043501;
O. Bertolami, P.J. Martins, \PR {\bf D61} (2000) 064007;
N. Banerjee, D. Pav\'on, \PR {\bf D63} (2001) 043504; \CQG {\bf 18}
(2001) 593;
A.A. Sen, S. Sen, S. Sethi, \PR {\bf D63} (2001) 107501;
A.A. Sen, S. Sen, \MPL {\bf A16} (2001) 1303;
 A. Albrecht, C. Skordis, \PRL 84 (2000) 2076.



\bibitem{quint2} Y. Fujii, \PR {\bf D61} (2000) 023504;
A. Masiero, M. Pietroni and F. Rosati, 
\PR {\bf D61} (2000) 023504;
 M.C. Bento, O. Bertolami, N.C.  Santos, \PR {\bf D65} (2002) 
067301.

\bibitem{Tegmark} H. Sandvik, M. Tegmark, M. Zaldarriaga, I. Waga,
 astro-ph/0212114.

\bibitem{Beca} L.M.G. Be\c{c}a, P.P Avelino, J.P.M. de Carvalho,
  C.J.A.P. Martins, astro-ph/0303564.

\bibitem{Bean} R. Bean, O. Dor\'e, astro-ph/0301308.

\bibitem{Supern} J.C. Fabris, S.B.V. Gon\c calves, P.E. de Souza,
astro-ph/0207430;
 P.P. Avelino, L.M.G. Be\c ca, J.P.M. de Carvalho,
 C.J.A.P. Martins, P. Pinto, astro-ph/0208528;
 A. Dev, J.S. Alcaniz, D. Jain, astro-ph/0209379;
V. Gorini, A. Kamenshchik, U. Moschella, astro-ph/0209395.


\bibitem{Makler} M. Makler, S.Q. de Oliveira, I. Waga, astro-ph/0209486.


\bibitem{Alcaniz} J.S. Alcaniz, D. Jain, A. Dev,  astro-ph/0210476.


\bibitem{Silva} P.T. Silva, O. Bertolami, astro-ph/0303353.



\bibitem{Bento4}  M. C. Bento, O. Bertolami and A. A. Sen, \PR {\bf D67 }
(2003) 063003.

\bibitem{Carturan} D. Carturan, F. Finelli, astro-ph/0211626.

\bibitem{Doran1} M. Doran, M. Lilley, J. Schwindt, C. Wetterich, 
astro-ph/0012139.


\bibitem{Doran2} M. Doran, M. Lilley, C. Wetterich, astro-ph/0105457.


\bibitem{Domenico} D. Di Domenico, C. Rubano, P. Scudellaro, 
astro-ph/0209357.

\bibitem{Barreiro} T. Barreiro, M.C. Bento, N.M.C. Santos and A.A. Sen,
 astro-ph/0303298.

\bibitem{Hu} W. Hu, M. Fukugita, M. Zaldarriaga, M. Tegmark,
\AJ {\bf 549} (2001) 669.


\bibitem{Boomerang} P. Bernardis et al., \AJ {\bf 564} (2002) 559.


\bibitem{WMAP} D.N. Spergel et al., astro-ph/0302207; L. Page et al., 
astro-ph/0302209. 


\bibitem{Hu1996} W. Hu, N. Sugiyama, \AJ {\bf 471} (1996) 30.

\bibitem{Reese} Reese et al., \AJ  {\bf 581} (2002) 53.


\bibitem{Mason} Mason et al., \AJ {\bf 555} (2001) L11.


\bibitem{Benoit} A. Benoit et al., astro-ph/0210306.

\bibitem{Amendola} L. Amendola, F. Finelli, C. Burigana, D. Carturan,
 astro-ph/0304325.

\end{thebibliography}
\end{document}